\documentclass[aps,prb,reprint,amssymb,amsmath,superscriptaddress]{revtex4-1}

\usepackage{graphicx}

\begin{document}

\title{Weak antilocalization in HgTe quantum well with inverted energy spectrum}

\author{G.~M.~Minkov}
\affiliation{Institute of Metal Physics RAS, 620990 Ekaterinburg,
Russia}

\affiliation{Institute of Natural Sciences, Ural Federal University,
620000 Ekaterinburg, Russia}

\author{A.~V.~Germanenko}

\author{O.~E.~Rut}
\affiliation{Institute of Natural Sciences, Ural Federal University,
620000 Ekaterinburg, Russia}

\author{A.~A.~Sherstobitov}
\affiliation{Institute of Metal Physics RAS, 620990 Ekaterinburg,
Russia}

\affiliation{Institute of Natural Sciences, Ural Federal University,
620000 Ekaterinburg, Russia}

\author{S.~A.~Dvoretski}

\affiliation{Institute of Semiconductor Physics RAS, 630090
Novosibirsk, Russia}

\author{N.~N.~Mikhailov}

\affiliation{Institute of Semiconductor Physics RAS, 630090
Novosibirsk, Russia}

\date{\today}

\begin{abstract}

The results of experimental study of the magnetoconductivity of 2D
electron gas caused by suppression of the interference quantum
correction in HgTe single quantum well heterostructure with the
inverted energy spectrum are presented. It is shown that only the
antilocalization magnetoconductivity  is observed at the relatively
high conductivity $\sigma> (20-30)\,G_0$, where $G_0= e^2/2\pi^2\hbar$.
The antilocalization correction demonstrates a crossover from
$0.5\,\ln{(\tau_\phi/\tau)}$ to $1.0\,\ln{(\tau_\phi/\tau)}$ behavior
with the increasing conductivity or decreasing temperature (here
$\tau_\phi$ and $\tau$ are the phase relaxation and transport
relaxation times, respectively). It is interpreted as a result of
crossover to the regime when the two chiral branches of the electron
energy spectrum contribute to the weak antilocalization independently.
At lower conductivity $\sigma<(20-30)\,G_0$, the magnetoconductivity
behaves itself analogously to that in usual 2D systems with the fast
spin relaxation: being negative in low magnetic field it becomes
positive in higher one. We have found that the temperature dependences
of the fitting parameter $\tau_\phi$ corresponding to the phase
relaxation time demonstrate reasonable behavior, close to $1/T$, over
the whole conductivity range from $5\,G_0$ up to $130\,G_0$. However,
the $\tau_\phi$ value remains practically independent of the
conductivity in distinction to the conventional 2D systems with the
simple energy spectrum, in which $\tau_\phi$ is enhanced with the
conductivity.

\end{abstract} 

\maketitle

\section{Introduction}
\label{sec:intr}

\begin{figure}
\includegraphics[width=0.8\linewidth,clip=true]{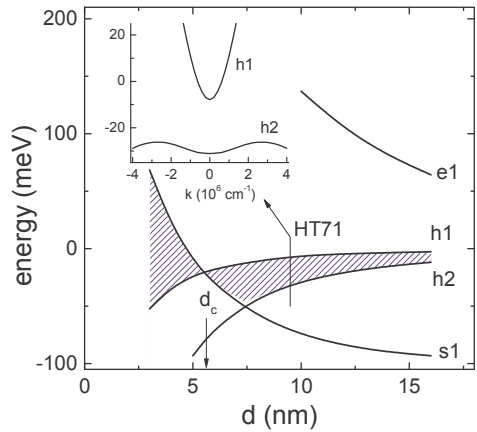}
\caption{(Color online) Energies of subbands at $k=0$ as a function of the HgTe
quantum well width for the Hg$_{0.35}$Cd$_{0.65}$Te/HgTe/Hg$_{0.35}$Cd$_{0.65}$Te
heterostructure calculated  in the $6\times 6$ $\bf{k\widehat{P}}$-model as described in
Ref.~\onlinecite{Larionova97} with parameters from Ref.~\onlinecite{Ortner02}.
The dashed area indicates the energy gap.
Inset shows the dispersion law for $d=9.5$~nm corresponding to the
structure HT71.}\label{f1}
\end{figure}

New type of two-dimensional (2D) systems, which energy spectrum  is
formed by the spin-orbit interaction has attracted considerable
interest during the  last decade. Just in these structures the
Dirac-like spectrum is realized. This leads to appearance of new and
modification of traditional dependences of kinetic phenomena on the
magnetic field, temperature, carriers density,  etc. The graphene,
\cite{Geim11,Novoselov11} topological insulators such as
Bi$_{1-x}$Sb$_x$, Bi$_2$Se$_3$, Bi$_2$Te$_3$,\cite{Hasan11,Hasan11-1}
quantum wells of gapless semiconductor HgTe\cite{Bernevig06,Buttner11}
belong to this type of the system. In the last-mentioned, the energy
spectrum can be tuned by changing of the quantum well width ($d$) from
inverted at $d>d_c$~nm to the normal one at $d<d_c$,\cite{Bernevig06}
where $d_c\simeq (6-7)$~nm for CdTe/HgTe/CdTe
heterostructure\cite{Gerchikov90} is the critical thickness of the HgTe
layer corresponding to the collapse of the energy gap (see
Fig.~\ref{f1}). Namely in the vicinity of $d_c$ the Dirac-like spectrum
is realized. Much progress in the growth technology of
Hg$_{1-x}$Cd$_x$Te/HgTe
heterostructures\cite{Goschenhofer98,Mikhailov06} gives possibility to
carry out the detailed studies of the transport phenomena in such a
type of 2D systems. Large number of the papers was devoted to studies
of high magnetic field transport,\cite{Zhang02,Zhang04,Gusev11} quantum
Hall effect,\cite{Ortner02,Gusev10} crossover from the electron to hole
conductivity in the gated structures.\cite{Koenig07,Kvon11} At the same
time, effects resulting from the quantum interference were investigated
in the sole experimental\cite{Olshanetsky10} and sole
theoretical\cite{Tkachov11} paper.

In this  work, we present the results of experimental study of the
interference quantum correction to the conductivity ($\delta\sigma$) in
the  HgTe single quantum well with the inverted energy spectrum. It is
shown that only the antilocalization magnetoconductivity  is observed
at relatively high conductivity $\sigma> (20-30)\,G_0$, where $G_0=
e^2/2\pi^2\hbar$. It has been found that the antilocalization
correction demonstrates a crossover from the regime when
$\delta\sigma=-\alpha\,\ln{(\tau_\phi/\tau)}$ with $\alpha=-0.5$ to
that with  $\alpha=-1.0$ ($\tau_\phi$ and $\tau$ are the phase
relaxation and transport relaxation times, respectively). It is
interpreted as a crossover to the regime when the two chiral branches
of the energy spectrum  contribute to the weak antilocalization
independently. At lower conductivity $\sigma<(20-30)\,G_0$, the
magnetoconductivity behaves itself analogously to that in usual 2D
systems with the fast spin relaxation. The magnetoconductivity curves
in this case are well fitted by the standard
expression.\cite{Hik80,Wit87} We  find  that the temperature
dependences of $\tau_\phi$, which value is found from the fit, is close
to $1/T$ law over the conductivity range $\sigma=(5-130)\,G_0$ as it
should be when the inelasticity of electron-electron ({\it e-e})
interaction is the main mechanism of the phase relaxation.\cite{AA85}
At the same time, the $\tau_\phi$ value remains practically independent
of the conductivity in contrast to the ordinary 2D systems, where
$\tau_\phi$ is roughly proportional to the conductivity.

\begin{figure}
\includegraphics[width=\linewidth,clip=true]{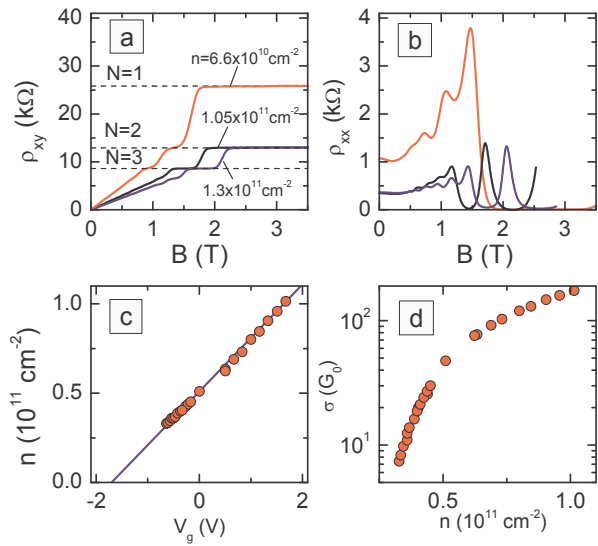}
\caption{(Color online) (a) and (b) -- The magnetic field dependences of $\rho_{xy}$  and $\rho_{xx}$,
respectively, measured at $T=1.35$~K for the different electron densities. (c) -- The gate voltage
dependences of the electron densities found as $n=-1/eR_H(0.1\,\texttt{T})$.
Symbols are the data, the line is the dependence $n=CV_g/|e|+5.1\times 10^{10}$ plotted with
$C=4.8$~nF/cm$^{-2}$ measured experimentally. (d) -- The conductivity measured at $B=0$ and $T=1.35$~K as
a function of the electron density.  }\label{f2}
\end{figure}

\section{Experimental}
\label{sec:expdet}

Our HgTe quantum wells  were realized on the basis of
HgTe/Hg$_{1-x}$Cd$_{x}$Te ($x=0.5-0.65$) heterostructure grown by means
of MBE on GaAs substrate with the (013) surface
orientation.\cite{Mikhailov06} The three heterostructures HT108, HT71,
and H922 with the nominal width of the quantum well equal to $9.0$~nm,
$9.5$~nm, and $10$~nm, respectively, were investigated. The samples
were mesa etched into standard Hall bars. To change and control the
electron density ($n$) in the quantum well, the field-effect
transistors was fabricated on the basis of the Hall bars with parylene
as an insulator and aluminium as gate electrode.  In some cases the
illumination was used to change the electron density in the quantum
well. The conductivity of the structure HT71 after the cooling down to
the liquid helium temperature was very low, $\sigma\simeq
10^{-2}\,G_0$, and enhanced up to $\simeq 130\,G_0$ with the help  of
illumination or application of the gate voltage. The electron density
in the structures HT108 and H922 at $V_g=0$ was about $4.5\times
10^{11}$~cm$^{-2}$ and $5.0\times 10^{11}$~cm$^{-2}$, respectively, and
decreased down to $(3-5)\times 10^{10}$~cm$^{-2}$ by the gate voltage.
The electron mobility at $n=1.0\times 10^{11}$~cm$^{-2}$ was $1.3\times
10^5$~cm$^2$/V\,s, $4.5\times 10^4$~cm$^2$/V\,s, and $4.5\times
10^4$~cm$^2$/V\,s in the structures HT71, HT108, and H922,
respectively. The main results were analogous for all the structures.
For this reason, the figures will represent the results obtained on the
structure HT71, except as otherwise noted.

\section{Results and discussion}
\label{sec:exp}

The magnetic field dependences of off-diagonal and diagonal component
of the resistivity tensor ($\rho_{xy}$ and $\rho_{xx}$) at some gate
voltages are presented in Fig.~\ref{f2}(a) and Fig.~\ref{f2}(b),
respectively. Well-resolved quantum Hall plateaus, both the even and
odd, are evident. The $V_g$ dependence of the electron density found as
$n=-1/eR_H(0.1\,\texttt{T})$, where $R_H$ is the Hall coefficient, and
the electron density dependence of the conductivity at $T=1.35$~K are
plotted in Fig.~\ref{f2}(c) and Fig.~\ref{f2}(d), respectively. As seen
$n$~vs~$V_g$ plot is close to the linear one and its slope corresponds
to the geometric capacity $C=4.8$~nF/cm$^{-2}$ measured on the same
sample.

\begin{figure}[b]
\includegraphics[width=\linewidth,clip=true]{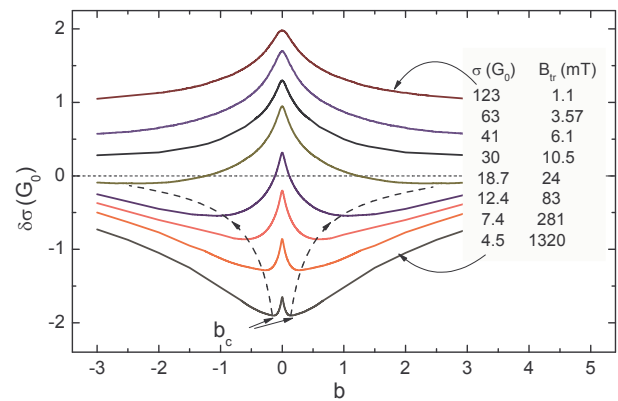}
\caption{(Color online) The dependences $\delta\sigma(b)=\Delta\sigma(b)+\delta\sigma(0)$ for
different conductivity values, $T=1.35$~K. The dashed arrows show the shift of the minimum
with the increasing conductivity.}\label{f3}
\end{figure}

Let us consider the low field magnetoconductivity presented in
Fig.~\ref{f3}. To compare the experimental curves measured at different
conductivity values, we plot them against the relative magnetic field
$b=B/B_{tr}$, where $B_{tr}=\hbar/2el^2$ with $l$ as the mean free path
is the characteristic magnetic field for the interference correction.
Furthermore, we shift the experimental dependences $\Delta\sigma(b)=
1/\rho_{xx}(b)- 1/\rho_{xx}(0)$  in the vertical direction on the value
of the interference correction at $B=0$, $\delta\sigma(0)$, found as
described in Section~\ref{sec:expc}. So, the plotted in Fig.~\ref{f3}
dependence is actually the $b$ dependence of the interference
correction $\delta\sigma(b)$. One can see that only the negative
(antilocalizing) magnetoconductivity is observed at high conductivity
$\sigma>(20-30)\,G_0$ over the whole magnetic field range up to $b=3$.
At lower conductivity values, the crossover to the positive
magnetoconductivity  is observed at the magnetic field labeled as
$b_c$. As clearly seen from the figure the lower the conductivity the
lower the value of $b_c$. At $\sigma<20\, G_0$ the value of  $b_c $
becomes less than unity.

\begin{figure}
\includegraphics[width=\linewidth,clip=true]{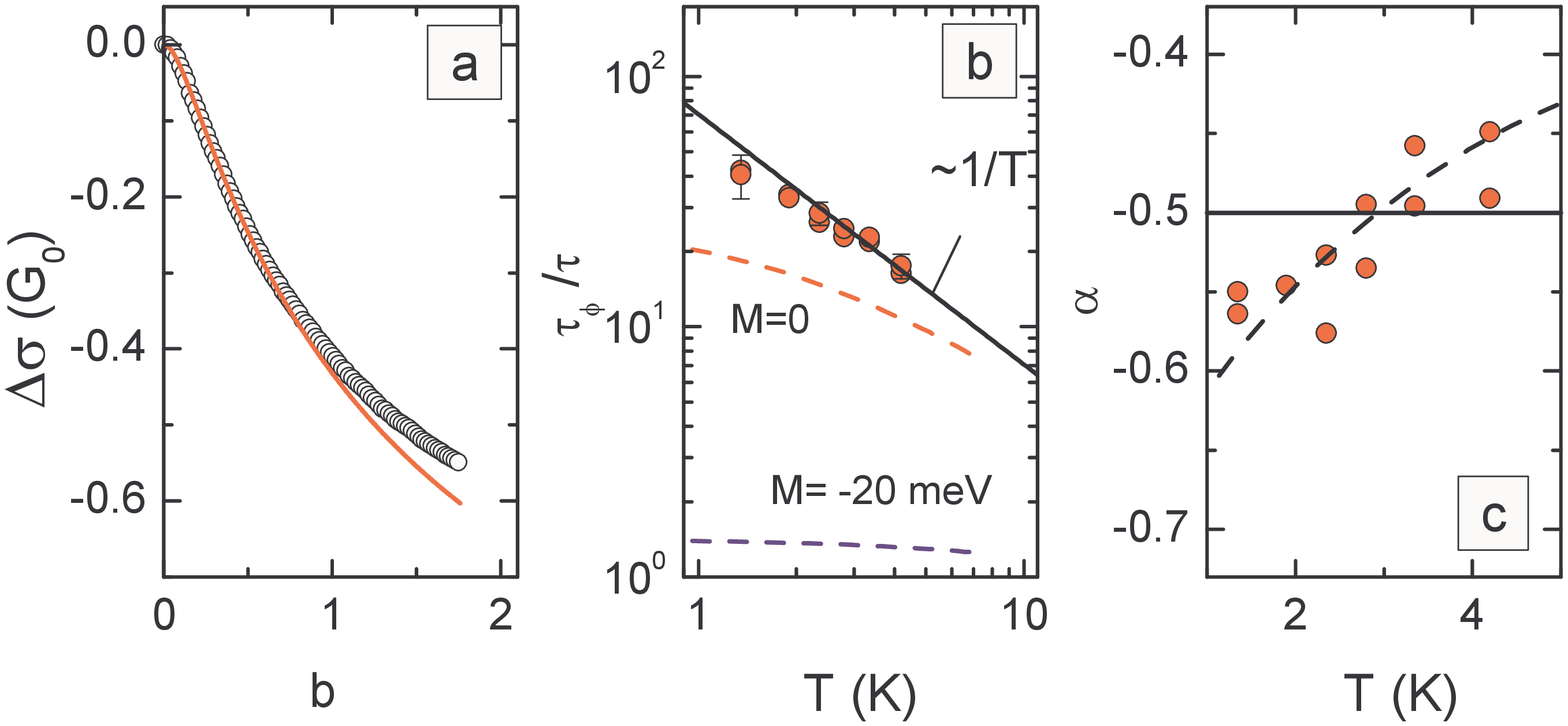}
\caption{(Color online) (a) -- The magnetic field dependence of
$\Delta\sigma$ for
$n=5.8\times 10^{10}$~cm$^{-2}$, $\sigma=63\,G_0$, $T=1.35$~K. Symbols are the data,
the curve is the results of the best fit by Eq.~(\ref{eq20}) carried out within the $b$ range
from $0$ to $0.3$.  (b) -- The temperature
dependence of the phase relaxation time found from the fit of the
magnetoconductivity curves. The symbols are the data. The dashed curves are calculated as described
in Section~\ref{sec:expa}. (c) -- The temperature dependence of the prefactor $\alpha$. The symbols are
the data found from the fit, the dashed line is provided as a guide to the eye, the solid line
is $\alpha=-0.5$.  }\label{f4}
\end{figure}

\subsection{Low-field positive magnetoconductivity, high conductivity $\sigma>20\,G_0$}
\label{sec:expa}

So far as we know the theory of the interference induced
magnetoconductivity for systems with complicated energy spectrum like
HgTe quantum wells is not developed yet. Therefore, our analysis will
lean upon the following qualitative consideration. As shown in
Ref.~\onlinecite{Bernevig06} the HgTe quantum wells have a
single-valley Dirac-like energy spectrum in the vicinity of the
critical thickness $d_c$. It consists of two branches of different
chirality (in what follows referred as $k^+$ and $k^-$), which are
degenerate in a symmetric quantum well or can be split-off due to
spin-orbit interaction in asymmetric case. The chiral fermions cannot
be backscattered and the magnetoconductivity in such a type of
heterostructures should demonstrate the antilocalization behavior. The
contribution to the interference correction coming from each branch is
positive and is equal to $0.5G_0\ln(\tau_\phi/\tau)$, $\tau_\phi\gg
\tau$. When the transition rate $1/\tau_\pm$ between the branches $k^+$
and $k^-$ is small as compared with the phase relaxation rate
$1/\tau_\phi$, the interference contributions to the conductivity from
these brunches are summarized and the total correction should be equal
to $(0.5+0.5)G_0\ln(\tau_\phi/\tau)$. In opposite case, when
$1/\tau_\pm\gg1/\tau_\phi$, the correction should be equal to  $0.5G_0
\ln(\tau_\phi/\tau)$. So, the interference correction at $B=0$ for
arbitrary relationship between $\tau_{\phi}$ and $\tau_\pm$ should be
equal to
\begin{equation}
 \delta\sigma=-\alpha\, G_0\ln{\left(\frac{\tau_\phi}{\tau}\right)},\,\,\,-1 \leqslant \alpha \leqslant -0.5.
 \label{eq10}
\end{equation}
The magnetoconductivity resulting from the suppression of the electron
interference by the magnetic field should be described by the standard
expression\cite{Hik80,Wit87} with same prefactor $\alpha$:
\begin{eqnarray}
\Delta\sigma(b)&=&\alpha\,G_0{\cal H}\left(\frac{\tau}{\tau_\phi},b\right), \nonumber \\
{\cal H}(x,y)&=&\psi\left(\frac{1}{2}+\frac{x}{y}\right)-\psi\left(\frac{1}{2}+\frac{1}{y}\right)-\ln{x},
\label{eq20}
\end{eqnarray}
where $\psi(x)$ is the digamma function. This equation with two fitting
parameters $\alpha$ and $\tau_\phi$ has been used for the quantitative
analysis of our magnetoconductivity curves.\footnote{Treating the
magnetoconductivity curves we have supposed that the electron effective
mass is independent of the electron density and is equal to
$0.022\,m_0$, which has been obtained from the Shubnikov-de Haas
experiments for $n=8\times 10^{10}$~cm$^{-2}$. Our calculations show
that the variation of the effective mass does not exceed (10-15)\%
within the actual range of the electron density from $3\times
10^{10}$~cm$^{-2}$ to $1.5\times 10^{11}$~cm$^{-2}$.} Because
Eq.~(\ref{eq20}) is correct for the diffusion regime, i.e., for $b\ll
1$,  the fitting range was restricted by the interval $0<b<0.3$. As an
example, the result of the fitting procedure made for $\sigma=63\,G_0$
is presented in Fig.~\ref{f4}(a). One can see that Eq.~(\ref{eq20})
well describes the run of experimental curve.

The temperature dependences of the fitting parameter $\tau_\phi/\tau$
as Fig.~\ref{f4}(b) shows is close to $1/T$ law that corresponds to the
inelasticity of {\it e-e} interaction as the the main mechanism of the
phase relaxation.\cite{AA85} Such the temperature dependence of
$\tau_\phi$ is observed over the entire conductivity range
$20\,G_0<\sigma<130\,G_0$.

Let us call our attention to the prefactor, which temperature
dependence is presented in Fig.~\ref{f4}(c). It is seen that the value
of $\alpha$ is close to $-0.5$ and it becomes more negative with the
decreasing temperature. The first-mentioned is indication of that the
inter-branch transition time $\tau_\pm$ is comparable with or less than
$\tau_\phi$. The fact that $|\alpha|$ is somewhat less than $0.5$ at
$T=4.2$~K can be explained by not rigorous fulfilment of the condition
$\tau_\phi\gg\tau$ under which Eq.~(\ref{eq20}) works. When this strong
inequality is violated, our fitting procedure gives the value of
$\tau_\phi$ close to the true one, whereas $\alpha$ occurs to be
reduced in magnitude.\cite{Min00-1} For the case presented in
Fig.~\ref{f4}, the ratio $\tau_\phi/\tau$ is about $15-20$ at $T=4.2$~K
that should results in reduction of $\alpha$ by a factor of about
$0.7$. The increase of $|\alpha|$ observed with the $T$ decrease may
result from enhancement of $\tau_\phi$ to $\tau_\pm$ ratio due to
increase  of the phase relaxation time with the decreasing temperature.
This fact together with that $\alpha$ becomes appreciably less than
$-0.5$ indicates that the system crosses over to the regime of
independent contributions of each chiral branch to the interference
correction.

Thus, the temperature dependences of both fitting parameters $\alpha$
and $\tau_\phi$ are sound.

\begin{figure}[t]
\includegraphics[width=\linewidth,clip=true]{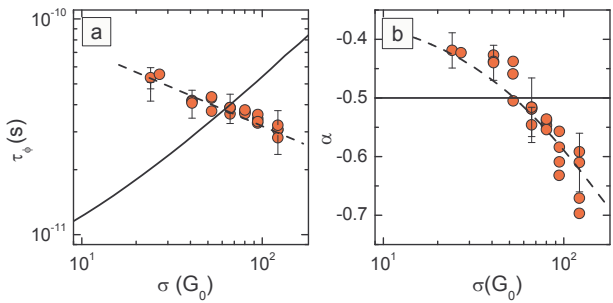}
\caption{(Color online) (a) -- The values of  $\tau_\phi$ (a) and $\alpha$ (b)
 plotted against the conductivity at $T=1.35$~K. The symbols are the data. The dashed lines are
 provided as a guide to the eye. The solid line in panel (a) is calculated
 in accordance with Ref.~\onlinecite{Zala01}. }\label{f5}
\end{figure}

We turn now to the conductivity dependence of $\tau_\phi$ and $\alpha$
shown in Fig.~\ref{f5}.  The surprising thing is that the parameter
$\tau_\phi$ does not increase with the increasing conductivity as
predicted theoretically\cite{Zala01} for the inelasticity of the {\it
e-e} interaction as the main mechanism of the phase relaxation [see
sold curve in Fig.~\ref{f5}(a)]. It should be also noted that the value
of $\tau_\phi$ found from the fit in HgTe quantum well at $\sigma
\simeq 20\,G_0$ is approximately $2.5-3$ times as large as the
theoretical value of the dephasing time. As Fig.~\ref{f5}(b)
demonstrates the prefactor $\alpha$ increases in magnitude with the
increasing conductivity. As discussed above such the behavior indicates
that the inter-branch transition time $\tau_\pm$ becomes larger than
the dephasing time $\tau_\phi$, and the 2D gas approaches the regime of
the two independent contributions of $k^+$ and $k^-$ states to the
interference quantum correction.

To the best of our knowledge there is the sole theoretical
paper\cite{Tkachov11} where the interference correction to the
conductivity in HgTe quantum well was studied. According to this paper
the value of the correction at $B=0$ is
\begin{eqnarray}
 \delta\sigma&=&-2\beta\,G_0\ln{\frac{\tau^{-1}}{\tau_{\cal M}^{-1}+\tau_\phi^{-1}}}, \label{eq30}\\
 \tau_{\cal M}^{-1}&=&\frac{2}{\tau} \frac{({\cal M}+{\cal B}k_F^2)^2}{{\cal A}^2k_F^2+({\cal M}+{\cal B}k_F^2)^2},\,\,\,k_F=\sqrt{2\pi n},
\label{eq40}
\end{eqnarray}
where ${\cal M}$ is the band gap at the Dirac point, ${\cal A}$ and
${\cal B}$ are the band parameters responsible for the linear and
quadratic parts of the energy spectrum, respectively,\cite{Bernevig06}
and $\beta>-1/2$ is the prefactor, which value depends on
$\tau/\tau_\phi$ and $\tau/\tau_{\cal M}$ so that $\beta\simeq -1/2$
when $\tau/\tau_\phi,\,\, \tau/\tau_{\cal M}\ll 1$ [see Eq.~(53) in
Ref.~\onlinecite{Tkachov11}]. Equation~(\ref{eq30}) is structurally
very similar to the  conventional expression, Eq.~(\ref{eq10}),
$1/\tau_\phi+1/\tau_{\cal M}$ stands instead of $1/\tau_\phi$ only. It
is natural to suppose that the magnetoconductivity could be described
by Eq.~(\ref{eq20}) with the same substitution. Under this assumption
the fit of the data by Eq.~(\ref{eq20}) should give
$1/\tau_\phi+1/\tau_{\cal M}$ instead of $1/\tau_\phi$. However,
following this line of attack we are not able to interpret our results.
To demonstrate this, we have depicted  the temperature dependences of
$\tau/\left(1/\tau_\phi+1/\tau_{\cal M}\right)$ calculated for the two
${\cal M}$ values, ${\cal M}=0$ and $-20$~meV, in Fig.~\ref{f4}(b). The
parameters ${\cal A}$ and ${\cal B}$ were equal to $380$~meV$\cdot$nm
and $850$~meV$\cdot$nm$^2$, respectively, in accordance with
Ref.~\onlinecite{Buttner11}, and $\tau/\tau_\phi$ varied with the
temperature as $\tau/\tau_\phi=0.014\,T$. One can see that contrary to
what observed experimentally, the calculated dependences demonstrate
very strong saturation with the temperature decrease, the saturation is
more pronounced for ${\cal M}=-20$~meV, which is more appropriate to
the structure investigated.

\subsection{Alternative sign magnetoconductivity, $\sigma<20\,G_0$}
\label{sec:expb}

\begin{figure}[t]
\includegraphics[width=\linewidth,clip=true]{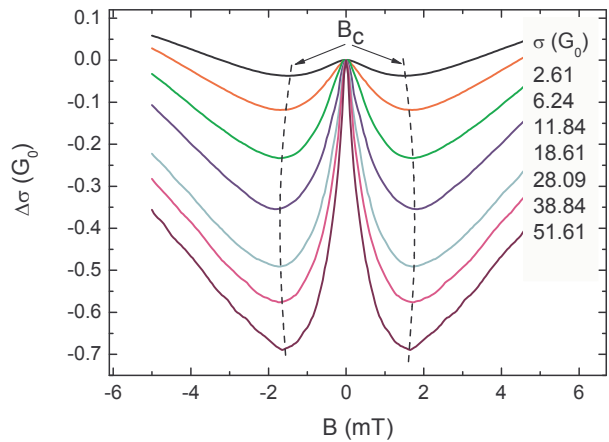}
\caption{(Color online) The magnetoconductivity plotted against the magnetic field
for different $\sigma$ at $T=1.35$~K. The dashed lines demonstrate that the minimum position is practically
independent of the conductivity. }\label{f6}
\end{figure}

Let us analyze the experimental results for the lower conductivity,
$\sigma\lesssim 20\,G_0$.  In this case, the crossover from the
negative magnetoconductivity to the positive one is observed at the
magnetic field, which value is lower than the transport magnetic field,
$b_c=B_c/B_{tr}\lesssim 1$ (see Fig.~\ref{f3}). At first sight
$\Delta\sigma(b)$ behaves much like the interference induced
magnetoconductivity observed in conventional 2D systems with the fast
spin relaxation (see, e.g., Ref.~\onlinecite{Knap96}). However, the
fact that the crossover field $B_c$ is practically independent of the
conductivity, $B_c=(15-20)$~mT, as clearly seen from Fig.~\ref{f6},
sets one thinking about other possible reasons responsible for the
crossover from the negative to positive magnetoconductivity at $B=B_c$.

In principle, the positive magnetoconductivity can result from the
contribution of the {\it e-e} interaction. It should depend on the
magnetic field due to the Zeeman splitting, which suppresses the
interaction contribution in the  triplet
channel.\cite{Cast84-2,Fin84,Raim90,Cast98,Zala01-2,Gor04,Min05-1} If
the value of the effective $\textsl{g}$-factor is sufficiently large,
the Zeeman splitting $\textsl{g}\mu_B B$, where $\mu_B$ stands for the
Bohr magneton,  can exceed the temperature already in relatively low
magnetic field, $B\sim 0.02$~T, and the positive magnetoconductivity
resulting from this effect can win the negative magnetoconductivity
caused by the quantum interference. Our measurements of the interaction
correction with the use of  the method suggested in
Ref.~\onlinecite{Min03-2} did not reveal any significant magnetic field
dependence of the interaction correction  in the actual magnetic-field
range. However, it must be admitted that the accuracy of this method at
so low field is not very high to assert it unambiguously.

Another possible  reason of the positive magnetoconductivity is
classical memory effects.\cite{Baskin78}  As shown in
Ref.~\onlinecite{Dmitriev01}, where the magnetotransport in the 2D
Lorenz gas is studied, these effects due to double scattering of an
electron on the same disk lead to a negative magnetoresistance even in
classically weak magnetic field.\cite{Dmitriev01,Cheianov04} However,
our estimations made according to Ref.~\onlinecite{Cheianov04} with the
use of typical for our case parameters show that the positive
magnetoconductivity due to the memory effect is more than one order of
magnitude less than the rise of the magnetoconductivity observed
experimentally at $B>B_c$.

So, the known mechanisms of positive magnetoconductivity cannot explain
our data. Therefore, despite the fact that the theory does not predict
alternative sign magnetoconductivity for systems with the Dirac-like
energy spectrum, let us analyze the data at $\sigma<20\,G_0$ under
assumption that both rising and ascending parts of the
magnetoconductivity curves in Fig.~\ref{f3} and Fig.~\ref{f6} result
from suppression of the interference correction in the magnetic field.
In what follows, we will compare the shape of the magnetoconductivity
curve with the well-known expressions employing the standard fitting
procedure.

For the case when the spin-orbit splitting of the energy spectrum is
cubic in quasimomentum the magnetoconductivity is described by the
Hikami-Larkin-Nagaoka expression\cite{Hik80}
\begin{eqnarray}
{\Delta\sigma(b)\over G_0}&=&\psi\left({1 \over 2}+{\tau\over
b}\left[{1\over\tau_\phi}+{1\over\tau_s}\right]\right)-\ln{\left({\tau\over
b}\left[{1\over\tau_\phi}+{1\over\tau_s}\right]\right)} \nonumber\\
&+&{1\over 2}\psi\left({1 \over 2}+{\tau\over
b}\left[{1\over\tau_\phi}+{2\over\tau_s}\right]\right)-{1\over
2}\ln{\left({\tau\over b}\left[{1\over\tau_\phi}+{2\over\tau_s}\right]\right)}\nonumber\\
&-& {1\over 2}\psi\left({1 \over 2}+{\tau\over
b}{1\over\tau_\phi}\right) +{1\over 2}\ln{\left({\tau\over
b}{1\over\tau_\phi}\right)}.
 \label{eq50}
\end{eqnarray}
Another case	is  the linear in quasimomentum splitting of the energy
spectrum. According to Ref.~\onlinecite{Iord94} (see also comments in
Ref.~\onlinecite{Min04-1}) $\Delta\sigma(b)$ in this case has the form
\begin{widetext}
\begin{eqnarray}
\label{eq60}
\frac{\Delta\sigma(b)}{G_0} &=& \frac{1}{2}\left[\sum_{n=1}^\infty \left\{ {3
\over n} - {3 a_n^2 + 2 a_n b_s - 1 - 2(2n+1) b_s \over
(a_n+b_s)a_{n-1}a_{n+1} - 2 b_s [(2n+1)a_n - 1]} \right\} - {1 \over
a_0} - \frac{2 a_0 + 1 +
b_s}{a_1(a_0+b_s) - 2 {b_s}} \right.\nonumber \\
&-& \left.2 \ln{(b_\phi + b_s)} - \ln{(b_\phi + 2 b_s)} - 3C -
S(b_\phi/b_s) - \Psi(1/2 + b_\phi) + \ln{b_\phi}\right],
\end{eqnarray}
\end{widetext}
where  $C \approx 0.57721$ is the Euler constant,
$b_\phi=b\,\tau/\tau_\phi$, $b_s=b\,\tau/\tau_s$, $a_n = n + 1/2 +
b_\phi + b_s$, and $S(b_\phi/b_s)$ is the $b$-independent function
\begin{equation}\label{eq70}
    S(x)={8 \over \sqrt{ 7 + 16x}} \left[\arctan{  { \sqrt{ 7 + 16x} \over 1 - 2
    x} }-\pi\Theta(1-2x)\right] \nonumber
\end{equation}
with $\Theta(y)$ as the Heaviside step function. Expressions
(\ref{eq50}) and (\ref{eq60}) have been derived under assumption that
the conductivity is very high, $\sigma \gg G_0$. When it is not the
case, one should regard for the second order quantum corrections. To
take them into account we have multiplied the right-hand side of both
expressions,  Eq.~(\ref{eq50}) and Eq.~(\ref{eq60}), by the factor
$1-2G_0/\sigma$ as it has been done for the case of slow spin
relaxation in Ref.~\onlinecite{Min04-2}.

\begin{figure}[t]
\includegraphics[width=\linewidth,clip=true]{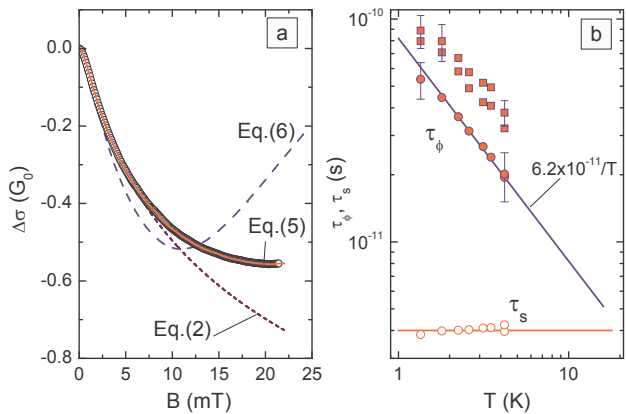}
\caption{(Color online) (a) -- The magnetoconductivity plotted as a function of the
magnetic field for  $\sigma =9.75\, G_0$ ($B_{tr}=62$~mT) at $T=1.35$~K.
The symbols are the data, the lines are the results of the fitting procedure by the different
expressions. (b) -- The temperature dependences of $\tau_\phi$ and $\tau_s$ obtained with the help
of Eq.~(\ref{eq50}) (circles) and Eq.~(\ref{eq20}) (squares). }\label{f7}
\end{figure}

In Fig.~\ref{f7}(a), we demonstrate the results of the fitting
procedure for $\sigma=9.75\, G_0$ carried out within the magnetic field
range $b=0-0.3$  with the use of $\tau_\phi$ and $\tau_s$ as the
fitting parameters. It is clearly seen that Eq.~(\ref{eq60}) does not
describe the experimental data, while Eq. (\ref{eq50}) gives very good
agreement. Such the agreement is observed at all the temperatures from
$1.3$~K to $4.2$~K over the conductivity interval from $5\, G_0$ to
$25\, G_0$. Therewith, the temperature dependences of the fitting
parameters $\tau_\phi$ and $\tau_s$ corresponding to the phase and spin
relaxation times, respectively, are reasonable. Again, $\tau_\phi$
demonstrates behavior close to $1/T$ law, while $\tau_s$ is temperature
independent within our accuracy  [see Fig.~\ref{f7}(b)] that is typical
for the degenerate gas of carriers.

\subsection{Overview of entire conductivity range, $\sigma=(5-130)\,G_0$}
\label{sec:expc}

\begin{figure}[b]
\includegraphics[width=\linewidth,clip=true]{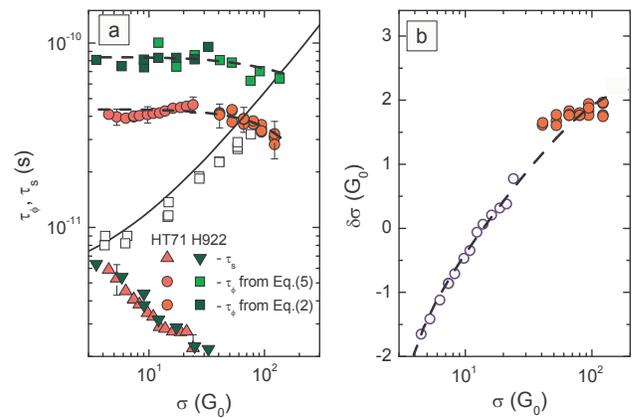}
\caption{(Color online) (a) -- The values of $\tau_\phi$ and $\tau_s$ plotted
against the conductivity for $T=1.35$~K for the structures HT71 and H922. The line is
calculated according to Ref.~\onlinecite{Zala01}. The open squares are the experimental results
obtained in Ref.~\onlinecite{Min04-2} on the sample H451 with GaAs/In$_{0.2}$Ga$_{0.8}$As/GaAs quantum well.
(b) -- The value of interference correction at zero magnetic field found from Eq.~(\ref{eq10}) (solid circles) and
Eq.~(\ref{eq70}) (open circles) as a function of the conductivity for $T=1.35$~K, structure
HT71. The dashed lines in both panels are provided as a guide to the eye.}\label{f8}
\end{figure}

Let us now inspect how the results obtained within the different
conductivity ranges dovetail into one another. The $\tau_\phi$ values
found for different conductivity regions as described above are graphed
as a function of the conductivity in Fig.~\ref{f8}(a).  The values of
the interference correction at $B=0$ calculated from Eq.~(\ref{eq10})
at high conductivity, $\sigma>20\,G_0$, and from the expression
\begin{equation}\label{eq70}
\frac{\delta\sigma(0)}{G_0}=-\frac{1}{2}\ln{\frac{\tau}{\tau_\phi}}+
\ln{\left(\frac{\tau}{\tau_\phi}+\frac{\tau}{\tau_s}\right)}+
\frac{1}{2}\ln{\left(\frac{\tau}{\tau_\phi}+\frac{2\tau}{\tau_s}\right)}
\end{equation}
at lower one are plotted in Fig.~\ref{f8}(b). It is seen that both
$\tau_\phi$~vs~$\sigma$ and $\delta\sigma$~vs~$\sigma$ dependences
found by the two different methods within the low- and
high-conductivity regions are matched well near $\sigma\simeq 30\,G_0$.
Namely these values of $\delta\sigma(0)$ have been used to shift the
experimental  dependences $\Delta\sigma(b)$ in the vertical direction
in Fig.~\ref{f3}. An essential feature evident from Fig.~\ref{f8}(a) is
that the fitting parameter $\tau_\phi$ does not practically depend on
the conductivity over the whole conductivity range from $\sigma\simeq
5\,G_0$ to $\sigma \simeq 130\, G_0$. Such the behavior is in
qualitative disagreement with that observed in conventional
A$_3$B$_5$-based  2D electron systems. For instance, the dephasing time
found experimentally in GaAs/In$_{0.2}$Ga$_{0.8}$As/GaAs single quantum
well heterostructures increases about five times over the same
conductivity range,\cite{Min04-2} that accords well with the
theoretical prediction\cite{Zala01} [see line and open squares in
Fig.~\ref{f8}(a)].

As already mentioned above the crossover to the positive
magnetoconductivity evident at $B_c$ can result from other mechanisms,
i.e., only the negative magnetoconductivity at $B<B_c$ is caused by
suppression of the electron interference, while another unknown
mechanism is responsible for the positive magnetoconductivity at
$B>B_c$. Then, Eq.~(\ref{eq20}) rather than  Eq.~(\ref{eq50}) should be
used to obtain the value of $\tau_\phi$ at low conductivity
$\sigma<20\,G_0$ as it took place for $\sigma>20\,G_0$, and the fitting
region should be restricted by the field lower than $B_c$. The result
of such a data treatment for $\sigma=9.75\,G_0$ ($B_{tr}=62$~mT) is
presented in Fig.~\ref{f7}(a) by the dotted line. The fit with the use
of $\tau_\phi$ and $\alpha$ as the fitting parameters has been made
within the magnetic field range $B=(0-1.5)$~mT. The prefactor $\alpha$
obtained from the fit is equal to $-0.33$, which is close to that
expected theoretically $\alpha\simeq -0.5\,(1-2G_0/\sigma)\simeq -0.4$.
The temperature dependence of $\tau_\phi$ is close to $1/T$, while the
value of $\tau_\phi$ is somewhat larger than that found with help of
Eq.(\ref{eq50}) [see Fig.~\ref{f7}(b)]. Analogous results are obtained
within whole low-conductivity range down to $\sigma=5\, G_0$, but, what
is more important, the use of Eq.~(\ref{eq20}) over the entire
conductivity range gives just the same result: $\tau_\phi$ dos not
increase with the increasing conductivity.

It is pertinent here to direct the reader's  attention to the results
obtained on the topological insulator Bi$_2$Se$_3$  and reported
recently in Ref.~\onlinecite{Hatke11}. The authors present the gate
voltage dependence of $B_\phi=\hbar/4eD\tau_\phi$ ($D$ is the diffusion
coefficient) found from the fit of the experimental magnetoconductivity
curves by Eq.~(\ref{eq20}). It is possible for some $V_g$ values to
convert these data to $\tau_\phi$~vs~$\sigma$ dependence. To do this
one should know the dependence of the effective mass ($m$) on the
electron density. If one believes that $m$ is constant in respect to
$n$, we obtain that $\tau_\phi$ in Bi$_2$Se$_3$ decreases with the
increasing conductivity within the range $\sigma = (50-500)\,G_0$.
However, if one naturally supposes that $m\propto \sqrt{n}$, that
corresponds to the linear Dirac-like spectrum, we obtain result
analogous to that presented above for HgTe quantum well: $\tau_\phi$ is
practically independent of $\sigma$.

Before closing this section let us list possible reasons for so
different behavior of $\tau_\phi$ with changing $\sigma$ in the 2D
systems with conventional parabolic energy spectrum and in the systems
with complicated spectrum like HgTe single quantum wells. First, the
parameters $\tau_\phi$ and $\tau_s$ found from the fit may not
correspond to the true phase and spin relaxation times despite the fact
that the standard expressions, Eq.~(\ref{eq20}) and Eq.~(\ref{eq50}),
fit the experimental magnetoconductivity curves rather well. Another
expression, which properly takes into account the peculiarities of the
energy spectrum and electron interference in the HgTe 2D systems should
be derived and used. Second, $\tau_\phi$ found from the fit is true or
close to that, but inelasticity of the {\it e-e} interaction in the
systems with complicated energy spectrum depends on the conductivity
really much weaker than in the conventional systems or there is more
effective additional mechanism of inelastic phase relaxation in the
structures under study that changes the dependence $\tau_\phi(\sigma)$
drastically. However, it remains unclear in the last case why the
dephasing time at low conductivity is five-to-ten times larger in HgTe
quantum well than that in conventional 2D systems  as Fig.~\ref{f8}(a)
illustrates.

\section{Conclusion}
We have studied the interference induced magnetoconductivity in single
quantum well of gapless semiconductor HgTe with the inverted energy
spectrum. It is shown that only the antilocalization
magnetoconductivity is observed at relatively high conductivity
$\sigma> (20-30)\,G_0$. The antilocalization correction demonstrates
the crossover from the $0.5\,\ln{(\tau_\phi/\tau)}$ to
$1.0\,\ln{(\tau_\phi/\tau)}$ behavior with the increasing conductivity
or decreasing temperature that is interpreted as a result of crossover
to the regime of independent contributions of the two chiral branches
to the weak antilocalization. At lower conductivity
$\sigma<(20-30)\,G_0$, the magnetoconductivity behaves itself
analogously to that in usual 2D systems with the fast spin relaxation.
It is  negative in low magnetic field and becomes positive in higher
one. We have found that the temperature dependences of the fitting
parameters $\tau_\phi$ corresponding to the phase relaxation times is
close to $1/T$ over the whole conductivity range $\sigma=(5-130)\,G_0$
that is typical for the dirty 2D systems at low temperature. However,
the $\tau_\phi$ value is practically independent of the conductivity
unlike the conventional 2D systems with the simple energy spectrum, in
which $\tau_\phi$ increases with the growing conductivity.

\section*{Acknowledgments}

We would like to thank I.~V.~Gornyi and I.~S.~Burmistrov for
illuminating discussions. This work has been supported in part by the
RFBR (Grant Nos.  10-02-91336 and 10-02-00481).

\end{document}